\documentclass[12pt,a4paper,final]{iopart} 

\usepackage{graphicx} 
\expandafter\let\csname equation*\endcsname=\relax 
\expandafter\let\csname endequation*\endcsname=\relax 
\usepackage{amsmath,amssymb,amsthm}
\usepackage{array}
\usepackage{xcolor}
\usepackage{CJKulem}
\usepackage{siunitx}
\usepackage{diagbox}
\DeclareSIUnit\Gauss{G}

\newcolumntype{P}[1]{>{\centering\arraybackslash}p{#1}}

\begin{document}

\title{Characterizing the Zeeman slowing force for $^{40}$Ca$^{19}$F molecules}




\author{P. Kaebert, M. Stepanova, T. Poll, M. Petzold, S. Xu\footnote{Corresponding Author}\addtocounter{footnote}{-1},  M. Siercke\footnotemark{} and S. Ospelkaus}			

\address{Institut f\"ur Quantenoptik, Leibniz Universit\"at Hannover, 30167~Hannover, Germany}	

\eads{s.xu@iqo.uni-hannover.de,siercke@iqo.uni-hannover.de}

\begin{abstract}
	In this paper we investigate the feasibility of Zeeman slowing calcium monofluoride (CaF) molecules originating from a cryogenic buffer gas cell. We measure the $A^2\Pi_{1/2} (v=0, J=\frac{1}{2}) - X^2\Sigma_{1/2} (v=0, N=1)$ hyperfine spectrum of CaF in the Paschen-Back regime and find excellent agreement with theory. We then investigate the scattering rate of the molecules in a molecular Zeeman slower by illuminating them with light from a \SI{10}{\milli\watt} broad repumper and a \SI{10}{\milli\watt} multi-frequency slowing laser. By comparing our results to theory we can calculate the photon scattering rate at higher powers, leading to a force profile for Zeeman slowing. We show results from a simple 1D simulation demonstrating that this force is both strong and narrow enough to lead to significant compression, and slowing of the molecules to trappable velocities.
	
\end{abstract}

\section{Introduction}

Owing to their internal degrees of freedom, molecular systems have the ability to probe fundamental physics and investigate states of matter dominated by long-range interactions. The ability to investigate these physical phenomena is directly linked to the number of molecules we can trap in experiments, as well as the temperatures to which we can cool them. To this end, experiments on laser cooling molecules have had incredible success, realizing magneto-optical traps \cite{barry_magneto-optical_2014, williams_characteristics_2017, collopy_3d_2018, mccarron_improved_2015, norrgard_submillikelvin_2016}, molasses \cite{anderegg_laser_2018, truppe_molecules_2017, kozyryev_sisyphus_2017, ding_sub-doppler_2020, cheuk__2018}, magnetic \cite{caldwell_deep_2019, williams_magnetic_2018, mccarron_magnetic_2018} and electrostatic traps \cite{prehn_optoelectrical_2016} as well as optical traps\cite{cheuk__2018, anderegg_optical_2019, anderegg_laser_2018}. Loading molecules into these traps generally requires a method of slowing a molecular beam down to velocities on the order of \SI{10}{\metre\per\second}. Currently, the most efficient method of laser slowing of molecules is chirped light slowing\,\cite{truppe_intense_2017}, where a fast moving molecular pulse, originating from a cryogenic buffer gas cell, is slowed down via photon scattering from a pulse of chirped light. The chirp of the laser is timed to compensate for the Doppler shift of the molecules as they are being slowed by the beam. Chirped light slowing is especially compatible with current buffer gas beam sources\,\cite{buffer_gas}, which deliver relatively short (in time) pulses of molecules.
In contrast, the predominant method of slowing down atoms using radiation pressure is the Zeeman slower\,\cite{Zeeman_atoms}. Here, instead of compensating for the Doppler shift with a chirp in the laser frequency, the Doppler shift is counteracted by a corresponding Zeeman shift from a magnetic field. By choosing the correct spacial variation of this magnetic field, a Zeeman slower is able to slow atoms to rest without the need to chirp the laser frequency, and without needing to know the time the atoms enter the slowing region. Zeeman slowing, unlike chirped light slowing, is thus a continuous slowing method, and would allow for the use of continuous molecular beam sources \cite{shaw_bright_2020}. This, coupled with the fact that the atoms are slowed down at a specific point in space, rather than a point in time, is the reason why atomic experiments opt for Zeeman slowing rather than chirped light slowing.

Zeeman slowing for molecules is not as straightforward as it is for atoms, due to the fact that molecular slowing schemes operate on a type-II transition ($J \rightarrow (J-1)$-transition). While this is necessary to achieve a quasi-closed cycling transition, it results in the molecules cycling between essentially all the ground and excited state Zeeman sublevels. Since Zeeman slowing relies on compensating the Doppler shift with a Zeeman shift, it requires all transitions the molecules undergo to have the same Zeeman shift, which is not the case for every Zeeman sublevel.

Nevertheless, Zeeman slowing of molecules was proposed to be possible \,\cite{petzold_zeeman_2018} by entering the Paschen-Back regime with a sufficiently high magnetic field. At such fields, the Zeeman shifts become highly regular, resembling a 3-level system. With the addition of one repump laser, this type-II Zeeman slower was shown to be capable of slowing Potassium atoms at efficiencies close to what a regular Zeeman slower is capable of \cite{petzold_type-ii_2018}. Zeeman slowing of an actual molecule however, has not yet been achieved.

In this paper, we measure the velocity dependence of the Zeeman slowing force for CaF molecules in magnetic fields between \SI{200}{\Gauss} and \SI{300}{\Gauss}. We first measure the strength and frequency of the individual hyperfine transitions of the $A^2\Pi_{1/2} (v=0, J=\frac{1}{2}) - X^2\Sigma_{1/2} (v=0, N=1)$ manifold at these fields by scanning a single frequency laser through the transition region. We then investigate the Zeeman slowing force by looking at the fluorescence of our molecular beam when it is illuminated by both a broad repump laser and a multi-frequency slowing laser. Unlike the schemes in \cite{petzold_type-ii_2018, petzold_zeeman_2018} and \cite{liang_improvements_2019} we do not tailor our slowing laser frequencies or polarization. Instead, the slowing laser contains both $\sigma^+$ and $\sigma^-$ polarizations and is frequency broadened by modulation of the laser current.

\section{Setup and Characterization of CaF transitions in the Paschen-Back regime}\label{sec:spectrum}

\begin{figure}
\begin{center}
	\includegraphics[scale=0.3]{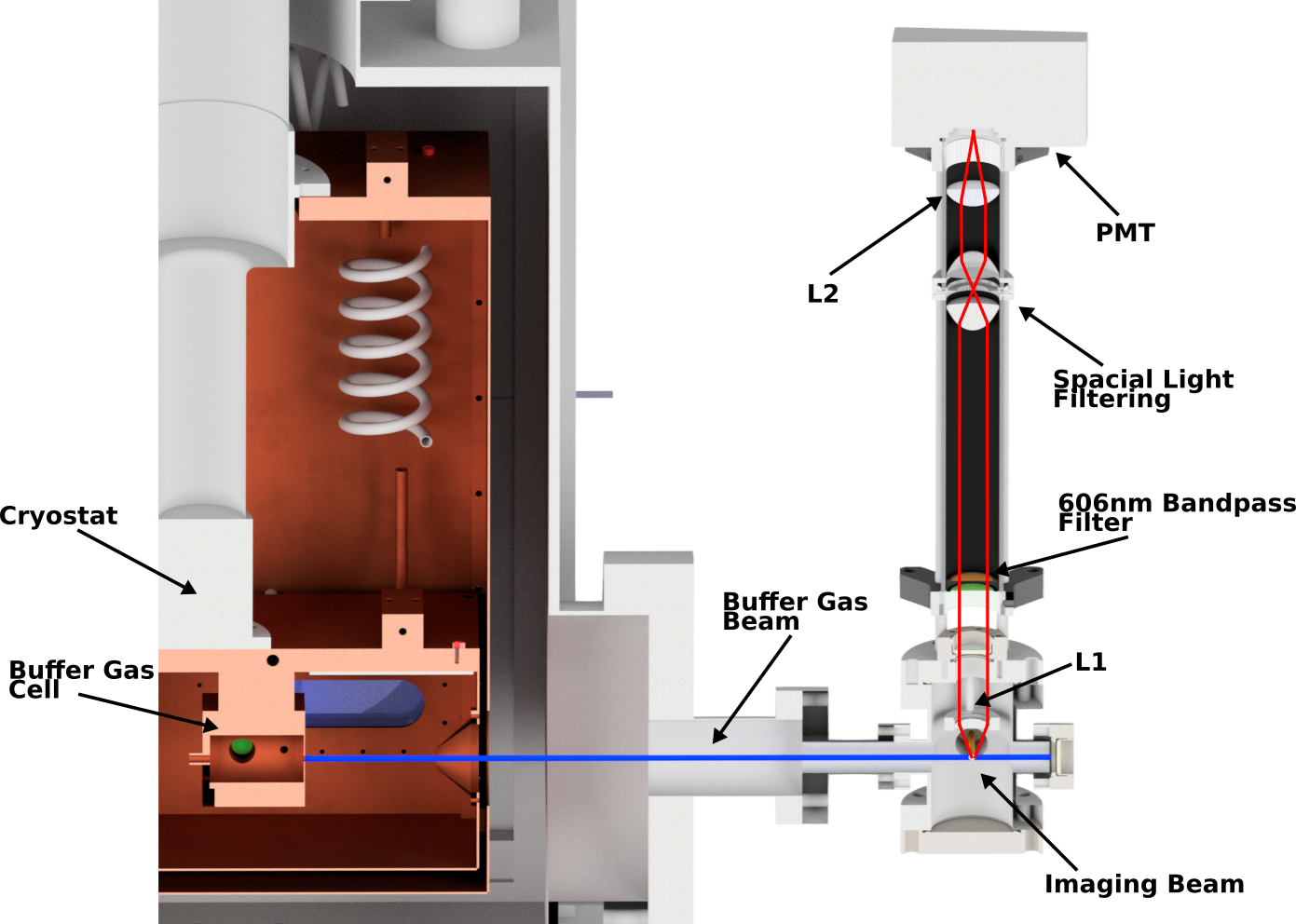}
	\caption{Setup of the experiment. Molecules are introduced into a cryogenic buffer gas cell via laser ablation of a $\mathrm{CaF_2}$ target, resulting in a beam of CaF molecules exiting the cell. \SI{33}{\centi\metre} away from the cell the CaF molecules enter the imaging region, where they are illuminated by either one single frequency, or two frequency broadened imaging beams. A pair of coils (not shown) can be used to generate a magnetic field along the imaging beam propagation direction, enabling spectroscopy of the molecules in magnetic fields up to \SI{300}{\Gauss}. }
	\label{fig:setup}
\end{center}
\end{figure}

The basic setup of our experiment is shown in figure \ref{fig:setup}. Our molecular beam is generated via laser ablation of a $\mathrm{CaF_2}$ target inside a cryogenic helium buffer gas cell. The beam exits the cell through a 3mm aperture and enters the detection region \SI{33}{\centi\metre} away from the exit. In the detection region, a \SI{2.0}{\milli\metre} wide by \SI{4.5}{\milli\metre} high elliptical beam with horizontal polarization intersects the molecular beam at a \SI{90}{\degree} angle, and the resulting fluorescence is collected onto a PMT running in photon counting mode. Using this setup we resolve lines with a separation as close as \SI[separate-uncertainty=true]{9 +- 1}{\mega\hertz}. A previous measurement of the $ A - X$ transition of CaF in high magnetic fields had a resolution of \SI{35}{\mega\hertz} (FWHM) and thus was not capable of resolving all individual hyperfine transitions \cite{devlin_measurements_2015}. Two magnetic field coils (not shown) surround the detection region, capable of producing a magnetic field of up to \SI{300}{\Gauss} along the propagation direction of the detection laser, allowing us to drive both $\sigma^+$ and $\sigma^-$ transitions with linear polarization, as was proposed in \cite{petzold_zeeman_2018}.

The detection laser consists of a laser diode at \SI{995}{\nano\metre}, amplified by a tapered amplifier and mixed with light out of a \SI{1550}{\nano\metre} fiber amplifier. The resulting \SI{606}{\nano\metre} light is stabilized on a scanning transfer cavity which is locked on a reference laser stabilized by a Potassium spectroscopy cell. The resulting frequency stability of the \SI{606}{\nano\metre}  light is $\approx\SI{1}{\mega\hertz}$, measured by beating two individual \SI{606}{\nano\metre} systems, locked on the same cavity, together. While this measurement does not account for any mutual drift of the two frequencies, our measurements on the molecular transitions did not show any evidence of such a drift.

\begin{figure}
\begin{center}
  \includegraphics[scale=1,viewport=0 0 360 577]{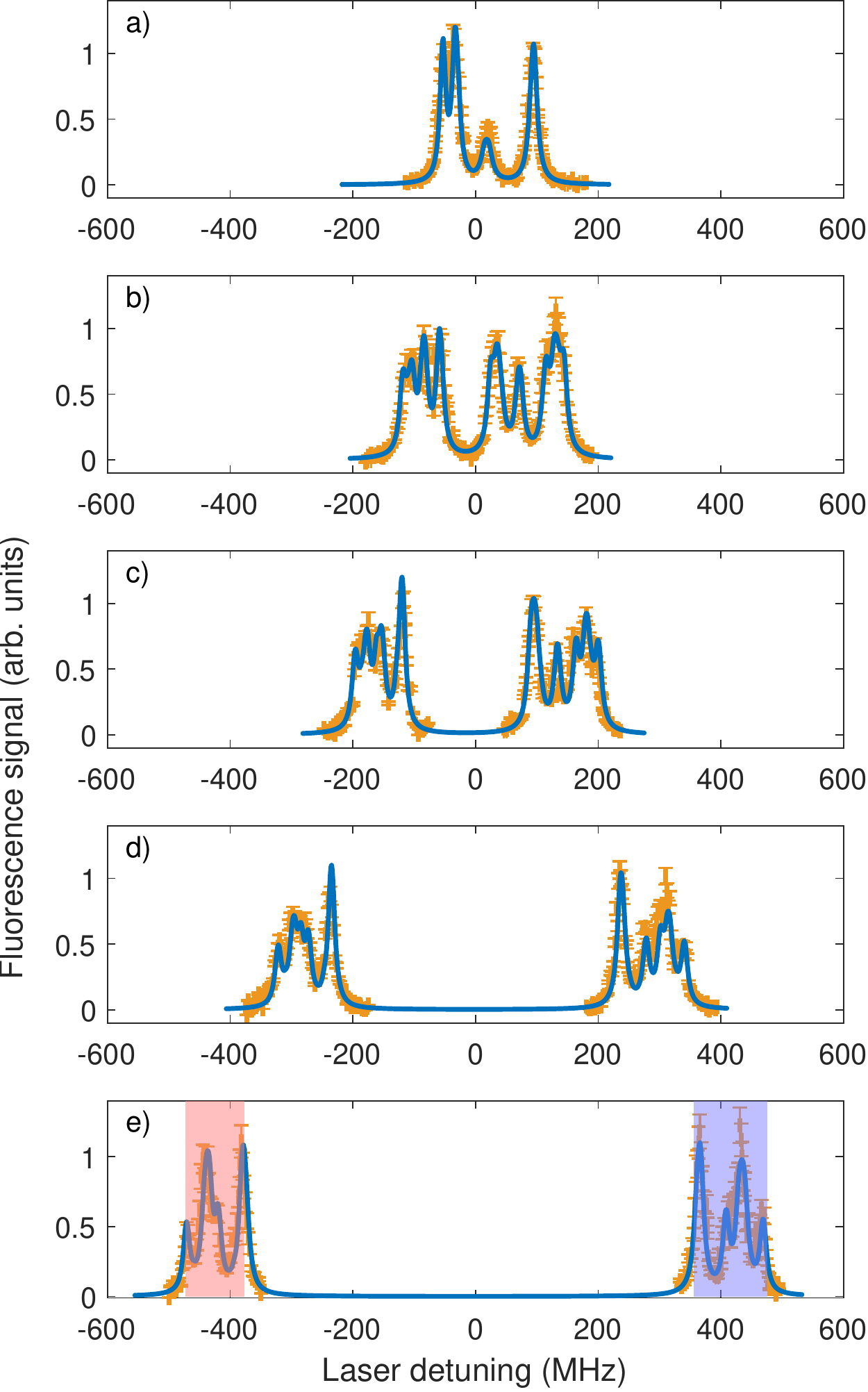}
  \caption{LIF Spectrum of CaF in fields of a) \SI{0}{\Gauss}, b) \SI{50}{\Gauss}, c) \SI{100}{\Gauss}, d) \SI{200}{\Gauss} and e) \SI{300}{\Gauss}. The blue line is the result of a rate equation model with a vertical scaling factor and a horizontal shift as the only free parameters. The left (red) and right (blue) rectangles in e) indicate respectively the widths of the slowing and repumper laser used to generate the Zeeman slower force.}
  \label{fig:spectrum}
\end{center}
\end{figure}

Figure \ref{fig:spectrum}\,a) shows the fluorescence signal of the molecules when no magnetic field is applied as the detection laser is scanned through the $A - X$ resonances. The laser power used in the figure is \SI{25}{\micro\watt} to avoid power broadening due to the absence of photon cycling in this measurement. Also shown in the figure is the result of a rate equation calculation (see \cite{tarbutt_magneto-optical_2015}) taking into account the interaction time between the molecules and the detection laser. The transition frequencies and strengths used by the program are calculated by converting the molecular states into the Hund's case (a) basis \cite{brown_rotational_2003} and calculating expectation values of the full Hamiltonian as well as the transition dipole matrix elements \cite{wall_lifetime_2008}. The details of this calculation are given in \ref{AppendixA}. Both the relative transition frequencies and the relative transition strengths measured in figure \ref{fig:spectrum} are in excellent agreement with theory. As such, the result of our rate equation calculation in figure \ref{fig:spectrum} is frequency shifted to properly fit the experimental data.

Figures \ref{fig:spectrum}b)-e) show the CaF fluorescence at magnetic fields of \SI{50}{\Gauss}, \SI{100}{\Gauss}, \SI{200}{\Gauss} and \SI{300}{\Gauss}. Again, both the relative positions and heights of the fluorescence peaks fit extremely well to theory.

\subsection{Measurement of the Zeeman slowing force}

\begin{figure}
\begin{center}
  \includegraphics[scale=1,viewport=0 0 357 363]{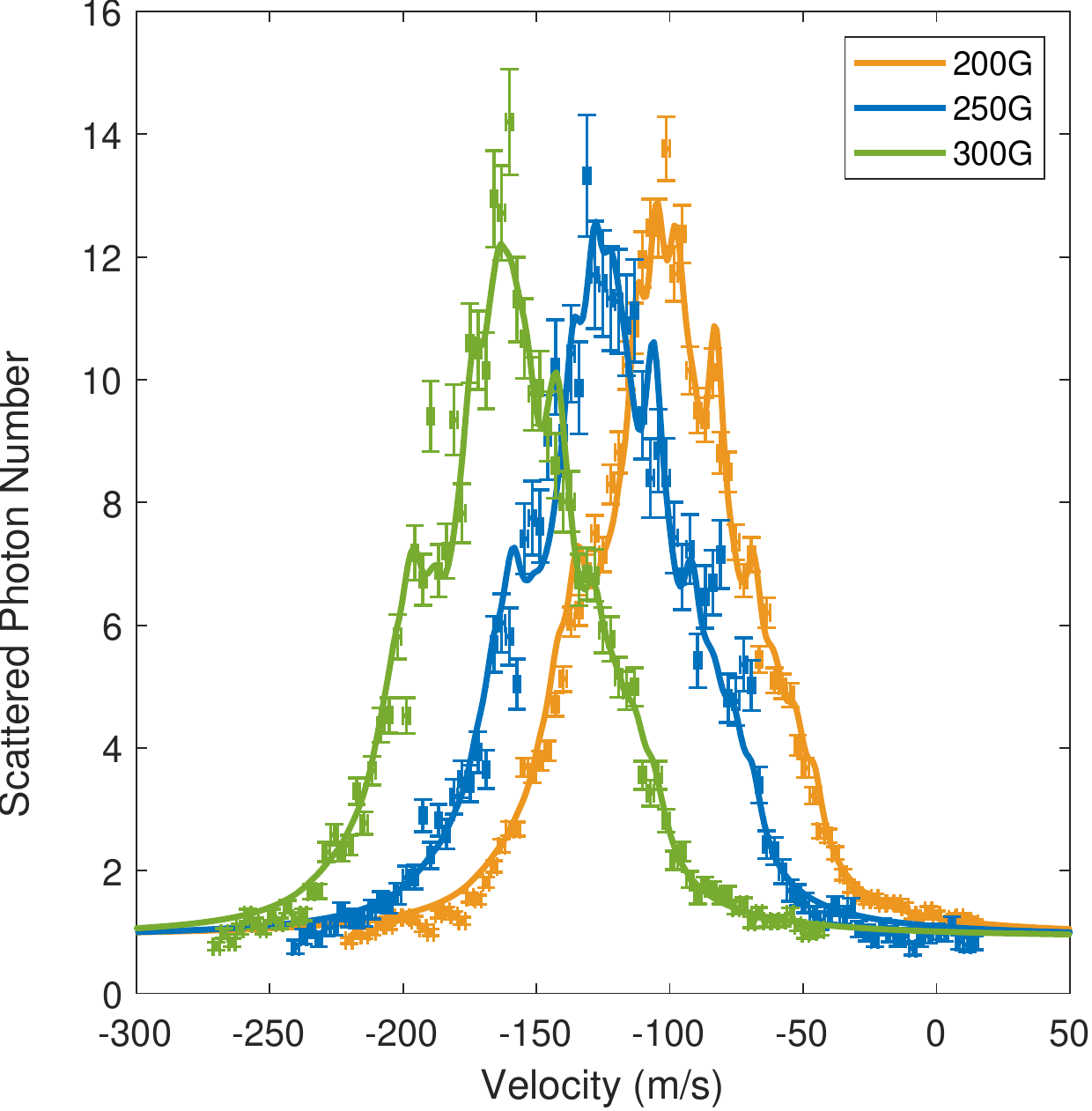}
  \caption{Scattered photon number vs slowing beam detuning of the molecules traversing the overlapped slowing and repumper beams at \SI{200}{\Gauss}, \SI{250}{\Gauss} and \SI{300}{\Gauss}. The vertical axis is calibrated by having enough repumper beam power to saturate the signal when the slowing beam is off-resonant and setting this off-resonant value equal to the theoretical photon scattering number of the rate equation model. The theory fits the data well with the molecular velocity as the only free parameter. The resulting velocity agrees reasonably well with our estimates for the beam velocity based on arrival times at the detector, giving a photon scattering rate of \SI[per-mode=reciprocal]{3.6e6}{\per\second}.}
  \label{fig:Zeeman_force}
\end{center}
\end{figure}

Using the knowledge that our theoretical prediction of the transition strengths and relative transition frequencies is accurate, we now turn to measure the strength and shape of the force a molecule would experience in a type-II Zeeman slower. To this end, we illuminate the molecules with both a repump laser, broadened by \SI{120}{\mega\hertz} and a multi-frequency slowing laser. Instead of carefully choosing the frequencies or polarizations of the slowing laser we measure the force when the slowing laser is frequency modulated to a width of \SI{96}{\mega\hertz}. At first glance it is not obvious that such a modulation can still provide a narrow enough force profile, since strong FM modulation will inherently result in a frequency spectrum with a tail. For the Zeeman slower force, however, we limit the modulation of the slowing laser to a modulation index of 3.9, resulting in a much sharper frequency cutoff than the one traditionally seen in white light slowing. Together with a modulation frequency of \SI{12}{\mega\hertz}, this gives our slowing laser its width of \SI{96}{\mega\hertz}, just enough to address all the transitions in the left cluster of peaks in figure \ref{fig:spectrum}e) indicated by the left (red) shaded region. The width of the broad repumper is indicated with the right (blue) shaded region. It should be noted, that in an actual Zeeman slower we would have to broaden the repumper much more, to cover all Zeeman and Doppler shifts the molecules experience, but our simulations show little dependence on the repumper width. In light of this we only modulate the repumper "enough" to reproduce the Zeeman slower force, thereby avoiding excessive scattering of repumper photons into our imaging system.
The fluorescence spectrum of the CaF molecules as they are illuminated by the broad repumper (\SI{10}{\milli\watt}) and slowing laser (\SI{10}{\milli\watt}) is shown in figure \ref{fig:Zeeman_force} for magnetic fields of \SI{200}{\Gauss}, \SI{250}{\Gauss} and \SI{300}{\Gauss}. Also plotted in the figure is the result of our rate equation model. At each magnetic field the frequency of the broad repump laser is held fixed, while the slowing laser is scanned just as it was for the single frequency measurements in section \ref{sec:spectrum} to simulate the force as a function of Doppler shift. As such, we plot the number of photons scattered in figure \ref{fig:Zeeman_force} as a function of velocity, rather than detuning of the laser. 

As was the case with figure \ref{fig:spectrum}, there is excellent agreement between the rate equation simulation and the experimental data. Two important conclusions can be drawn from this agreement. Firstly, no coherent dark states are formed during the application of the slowing force. Since our model would not be able to capture these dark state, their signature would be a decrease in the photon scattering rate compared to what our model predicts. While the schemes in \cite{petzold_type-ii_2018, petzold_zeeman_2018} and \cite{liang_improvements_2019} attempt to circumvent these dark states via carefully choosing the slowing frequencies or polarizations, our results demonstrate that simple frequency modulation at \SI{12}{\mega\hertz} is good enough to destabilize them.
Secondly the good agreement between the model and the experiment allows us to extract the photon scattering rate, and thus the magnitude of the Zeeman slowing force acting upon the molecules. To do this, we operate in a regime where, in the absence of scattering from the slowing laser, the repumper saturates the photon scattering rate. This results in the background signal in figure \ref{fig:Zeeman_force}, which, due to saturation, is almost independent of the laser power and the interaction time with the molecules. Due to this independence on experimental parameters we therefore know how many photons each molecule scatters when the slowing beam is far off-resonant, which calibrates the vertical axis in figure \ref{fig:Zeeman_force}. From the ratio between the on-resonant and the off-resonant photon scattering number we can then extract the interaction time of the molecules with the laser beams to be \SI{4.4}{\micro\second}. This time is consistent with our estimated laser beam diameter and buffer gas beam velocity (based on arrival time at the detector), and thus gives us a photon scattering rate of \SI[per-mode=reciprocal]{3.6e6}{\per\second} for our parameters.


\subsection{Zeeman Slowing Theory}
\begin{figure}
\begin{center}
	\includegraphics[scale=1,viewport=0 0 354 281]{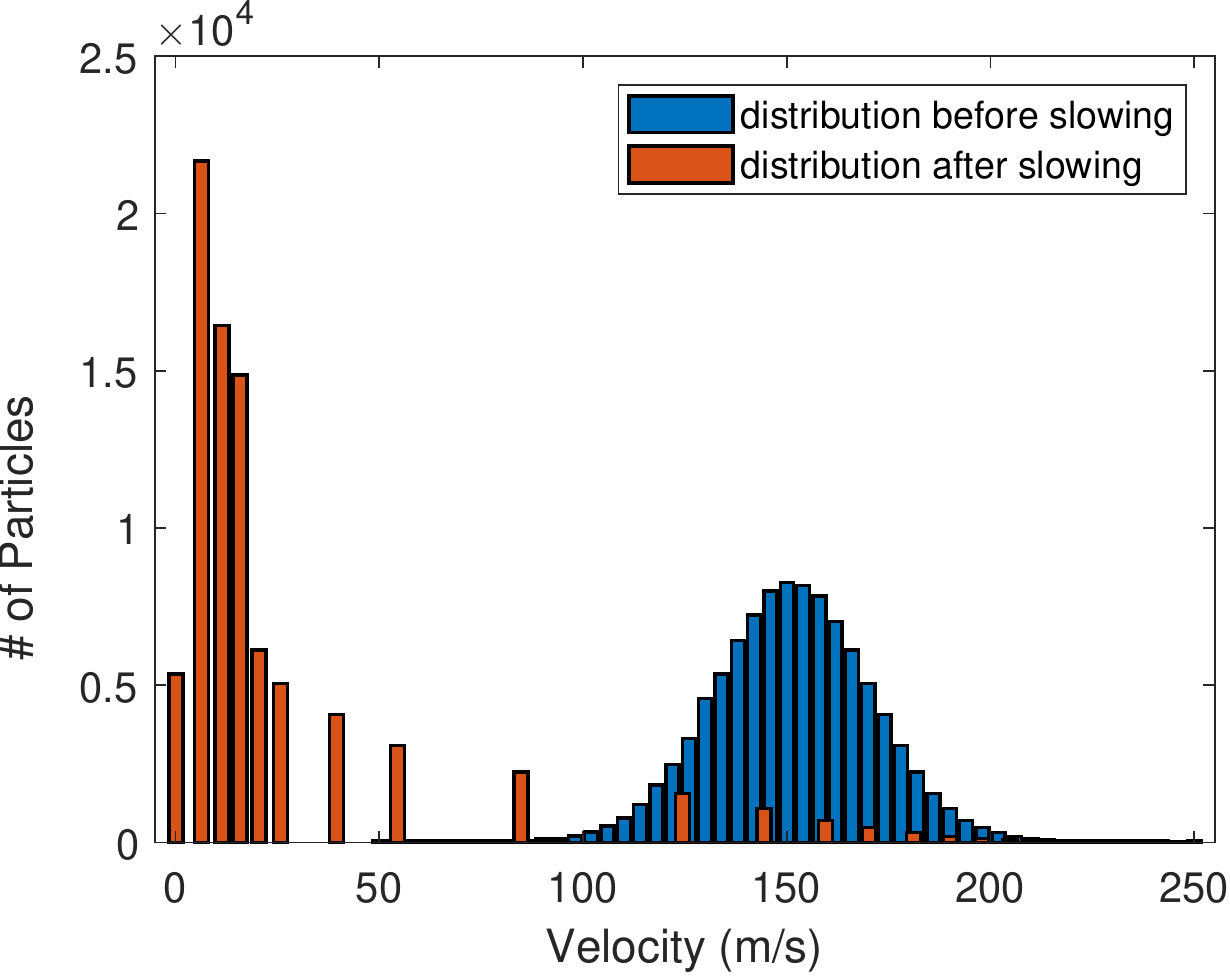}
	\caption{Simulated velocity distribution before and after Zeeman slowing. For the chosen parameters see text.}
	\label{fig:Zeeman_slower}
\end{center}
\end{figure}

With the data from figures \ref{fig:spectrum} and \ref{fig:Zeeman_force} confirming the accuracy of our rate equation calculation, we now know the photon scattering rate as a function of Doppler shift and magnetic field, and can thus simulate if the resulting force profiles are narrow and large enough to allow for efficient Zeeman slowing. It should be noted, however, that the interaction times and powers for the measurement in \ref{fig:Zeeman_force} were chosen such that decay to the $v=1$ ground state would be small. The necessity to eventually cycle molecules that have dropped to $v=1$ back to the vibrational ground state comes with the consequence that the slowing force is essentially halved \cite{petzold_zeeman_2018}. Figure \ref{fig:Zeeman_slower} shows the result of a simple 1D simulation of the molecular velocity before and after Zeeman slowing, taking this factor of 2 reduction in the force into account. The initial velocity distribution in the theory is centered on a velocity of \SI{150}{\metre\per\second} with a width of \SI{50}{\metre\per\second}. The magnetic field assumed in the simulation follows the standard square root Zeeman slower profile from
\SI{400}{\Gauss} to \SI{600}{\Gauss} \cite{atomic_zeeman}. The simulation assumes a \SI{995}{\milli\watt\per\centi\metre\squared} repump laser modulated by \SI{12}{\mega\hertz} to a width of \SI{960}{\mega\hertz}, and a \SI{240}{\milli\watt\per\centi\metre\squared} slowing laser modulated by \SI{15}{\mega\hertz}. The center frequency of the slowing laser is chosen to slow molecules with velocities up to \SI{180}{\metre\per\second}. The length of the slowing region is \SI{0.6}{\metre}. As can be seen in the figure, the initially broad, fast molecular velocity distribution is compressed and shifted down to \SI{15}{\metre\per\second} by the Zeeman slowing force, indicating that the profiles measured in \ref{fig:Zeeman_force} are indeed narrow enough, and provide enough photon scattering, to efficiently compress and slow the velocity distribution of the molecular beam.

\section{Conclusion}
We have measured the hyperfine spectrum of the $A^2\Pi_{1/2}(v=0, J=\frac{1}{2}) - X^2\Sigma_{1/2}(v=0, N=1)$ transition in CaF for magnetic fields up to 300G. We have found excellent agreement both in position and height of the transition peaks compared to a rate equation model, with a measurement resolution of \SI[separate-uncertainty=true]{9 +- 1}{\mega\hertz}. To measure the force a molecule would experience inside a type-II Zeeman slower we have measured the LIF signal of the molecules when subjected to \SI{10}{\milli\watt} of light from a \SI{120}{\mega\hertz} broad repumper laser and a \SI{96}{\mega\hertz} broad slowing laser, addressing the six lower, and six upper ground state hyperfine levels respectively. We again found excellent agreement with theory, measuring a peak photon scattering number of $16$ photons in the \SI{4.4}{\micro\second} the molecules spend inside the laser beams, demonstrating that we are already cycling between ground state sublevels at these modest interaction times and powers. We modeled 1D Zeeman slowing of CaF using the force profiles given by the rate equation model, including effects from $v=1$ repumping, and found effective slowing and compression of molecules down to 
\SI{15}{\metre\per\second}, where they can be captured by a MOT. Our findings directly show that efficient Zeeman slowing is possible with reasonable laser powers, and that a simple modulation of the slowing beam destabilizes any dark states while keeping the force profile narrow enough for velocity compression. With all the building blocks experimentally proven, there seem to be no obstacles on the road to realizing a Zeeman slower for molecules.

\section*{Acknowledgement}

 P. K., M. St. and M. S. thank the DFG for financial support through RTG 1991. We gratefully acknowledge financial support through  Germany’s Excellence Strategy – EXC-2123/1 QuantumFrontiers.

\newpage
\appendix
\section{Calculating the Eigenenergies and Transition Rates of CaF} \label{AppendixA}

Our approach to calculating the energies and eigenstates of CaF in various magnetic fields is essentially taken from \cite{brown_rotational_2003}. We summarize our method here for clarity as well as to highlight a discrepancy in \cite{brown_rotational_2003}. The calculation of the transition matrix elements is based off of Appendix A in \cite{wall_lifetime_2008}. For a great complementary discussion of the various parts of the Hamiltonian, see \cite{fitch_laser_2021}
The Molecular Hamiltonian for a particular electronic state can be divided up into different parts:

\begin{equation}\label{Hamiltonian}
\hat{H}=\hat{H}_{so}+\hat{H}_{ss}+\hat{H}_{rot}+\hat{H}_{sr}+\hat{H}_{LD}+\hat{H}_{F}+\hat{H}_{dip}+\hat{H}_{z}.
\end{equation}
Here, $\hat{H}_{so}$ is the spin-orbit interaction, $\hat{H}_{ss}$ is the spin-spin interaction (zero in our case), $\hat{H}_{rot}$ is the rotational part, $\hat{H}_{sr}$ describes the spin-rotation interaction, $\hat{H}_{LD}$ the lambda doubling, $\hat{H}_{F}$ and  $\hat{H}_{dip}$ are hyperfine terms describing the Fermi contact and magnetic dipolar interaction respectively, and $\hat{H}_{z}$ describes the Zeeman interaction of the molecule. For $\Sigma$ states $\hat{H}_{LD}$=0, while $\hat{H}_{LD}$ takes on the role of $\hat{H}_{sr}$ for $\Pi$ states, and $\hat{H}_{sr}$=0. In order to write down the specific form of each of these terms in the Hamiltonian we will need to choose a basis. Unfortunately, for the case of laser-coolable molecules, the ground and excited states are best described with a different set of quantum numbers. The excited, $A\Pi_{1/2}$ State is quasi-diagonal in a Hunds case (a) basis $\rvert\Lambda,S,\Sigma,\Omega,J,m_j,I,m_i\rangle$ while the $X\Sigma_{1/2}$ ground state is best described in Hunds case (b) $\rvert \Lambda,N,S,J,m_j,I,m_i\rangle$. Since the choice of basis is arbitrary we choose to work in the Hunds case (a) basis. Following section 9.4.2c) in \cite{brown_rotational_2003} we write the explicit form of the rotational and spin-orbit terms as

\begin{equation}
\begin{split}
\langle \Lambda&,S,\Sigma,\Omega,J,m_j,I,m_i \lvert \hat{H}_{rot}+\hat{H}_{so} \rvert \Lambda,S,\Sigma',\Omega',J,m_j,I,m_i \rangle \\
&=\delta_{\Omega,\Omega'}\delta_{\Sigma,\Sigma'}\lbrack B_0\lbrace J(J+1)+S(S+1)-2\Omega\Sigma-\Lambda^2\rbrace+A\Lambda\Sigma\rbrack\\
&-2B_0\sum_{q=\pm1}^{}(-1)^{J+S-\Omega-\Sigma}
\left(\begin{array}{ccc}
J & 1 & J\\
-\Omega & q & \Omega'
\end{array}\right)
\left(\begin{array}{ccc}
S & 1 & S\\
-\Sigma & q & \Sigma'
\end{array}\right)\\
&\times\sqrt{J(J+1)(2J+1)S(S+1)(2S+1)},
\end{split}
\end{equation}
where $B_0$ is the rotational constant of the $v=0$ vibrational state.

From \cite{asensio_ramos_theory_2006} we find
\begin{equation}
\begin{split}
\langle \Lambda&,S,\Sigma,\Omega,J,m_j,I,m_i \lvert \hat{H}_{sr}\rvert \Lambda,S,\Sigma',\Omega',J,m_j,I,m_i \rangle \\
&=\delta_{\Sigma,\Sigma'}\delta_{\Omega,\Omega'}\gamma[\Omega\Sigma-S(S+1)]\\
&+\gamma\sum_{q=\pm1}^{}(-1)^{J+S-\Omega-\Sigma} \left(\begin{array}{ccc}
J & 1 & J\\
-\Omega & q & \Omega'
\end{array}\right)
\left(\begin{array}{ccc}
S & 1 & S\\
-\Sigma & q & \Sigma'
\end{array}\right)\\
&\times\sqrt{J(J+1)(2J+1)S(S+1)(2S+1)},
\end{split}
\end{equation}
with $\gamma$ the spin-rotation constant.

From equation 9.66 in \cite{brown_rotational_2003}
\begin{equation}
\begin{split}
\langle \Lambda&,S,\Sigma,\Omega,J,m_j,I,m_i \lvert \hat{H}_{LD}\rvert \Lambda',S,\Sigma',\Omega',J,m_j,I,m_i \rangle \\
&=\sum_{r=\pm1}\delta_{\Lambda',\Lambda\pm2}\lbrace\delta_{\Sigma,\Sigma'}
\frac{q}{2\sqrt{6}}(-1)^{J-\Omega}\left(\begin{array}{ccc}
J & 2 & J\\
-\Omega & -2r & \Omega'
\end{array}\right)\\
&\times\sqrt{(2J-1)(2J)(2J+1)(2J+2)(2J+3)}+(p+2q)(-1)^{J+S-\Omega-\Sigma}\\
&\times\left(\begin{array}{ccc}
J & 1 & J\\
-\Omega & -r & \Omega'
\end{array}\right)\left(\begin{array}{ccc}
S & 1 & S\\
-\Sigma & r & \Sigma'
\end{array}\right)\\
&\times\sqrt{J(J+1)(2J+1)S(S+1)(2S+1)}
\rbrace.
\end{split}
\end{equation}
Here, $p$ and $q$ are the lambda doubling parameters of the state.

The two hyperfine terms are given in \cite{brown_rotational_2003} in the  $\lvert \Lambda,S,\Sigma,\Omega,J,I,F,m_f\rangle$ basis in equations (9.47) and (9.49) (although there is a factor 1/2 discrepancy between 7.156 and 9.61) as

\begin{equation}
\begin{split}
\langle \Lambda&,S,\Sigma,\Omega,J,I,F,m_f\rvert \hat{H}_F\lvert \Lambda,S,\Sigma',\Omega',J',I,F,m_f\rangle\\
&=b_F\sum_{q=-1}^{1}(-1)^{I+J'+F+S-\Sigma+J-\Omega}\\
&\times\sqrt{I(I+1)(2I+1)(2J+1)(2J'+1)S(S+1)(2S+1)}\\
&\times\left\lbrace\begin{array}{ccc}
J' & I & F\\
I & J & 1
\end{array}\right\rbrace
\left(\begin{array}{ccc}
J & 1 & J'\\
-\Omega & q & \Omega'
\end{array}\right)
\left(\begin{array}{ccc}
S & 1 & S\\
-\Sigma & q & \Sigma'
\end{array}\right),
\end{split}
\end{equation} 
and
\begin{equation}
\begin{split}
\langle \Lambda&,S,\Sigma,\Omega,J,I,F,m_f\rvert \hat{H}_{dip}\lvert \Lambda,S,\Sigma',\Omega',J',I,F,m_f\rangle\\
&=\sqrt{30}t(-1)^{I+J'+F+S+q-\Sigma+J-\Omega}\\
&\times\sqrt{I(I+1)(2I+1)(2J+1)(2J'+1)S(S+1)(2S+1)}
\left\lbrace\begin{array}{ccc}
J' & I & F\\
I & J & 1
\end{array}\right\rbrace\\
&\left(\begin{array}{ccc}
J & 1 & J'\\
-\Omega & q & \Omega'
\end{array}\right)
\left(\begin{array}{ccc}
1 & 2 & 1\\
q & 0 & -q
\end{array}\right)
\left(\begin{array}{ccc}
S & 1 & S\\
-\Sigma & q & \Sigma'
\end{array}\right).
\end{split}
\end{equation}
The constants $b_F$ and $t$ are related to the $b$ and $c$ parameters of Frosch and Foley \cite{frosch_magnetic_1952} by $b_F=b+1/3c$ and $t=c/3$.
To write the hyperfine terms in our uncoupled $\rvert\Lambda,S,\Sigma,\Omega,J,m_j,I,m_i\rangle$ basis we write equation (5.79) in \cite{brown_rotational_2003} explicitly as
\begin{equation}
\begin{split}
\rvert\Lambda&,S,\Sigma,\Omega,J,m_j,I,m_i\rangle=\sum_{F=|J-I|}^{J+I}\sum_{m_f=-F}^{F}(-1)^{-J+I-m_f}
\sqrt{2F+1}\\
&\times\left(\begin{array}{ccc}
J & I & F\\
m_j & m_i & -m_f
\end{array}\right)\lvert \Lambda,S,\Sigma',\Omega',J',I,F,m_f\rangle
\end{split}
\end{equation}
and compute the matrix elements of $\hat{H}_F$ and $\hat{H}_{dip}$ in our uncoupled basis as sums of the matrix elements in the coupled basis .

Finally, we write the Zeeman part of the Hamiltonian as in \cite{brown_rotational_2003} equation (9.71), neglecting the nuclear and rotational g-factors (which are $\sim1000$ times smaller than the electronic ones) as

\begin{equation}
\begin{split}
\langle \Lambda&,S,\Sigma,\Omega,J,m_j,I,m_i \lvert \hat{H}_{z}\rvert \Lambda',S,\Sigma',\Omega',J,m_j,I,m_i \rangle \\
&=\mu_B B_z\delta_{\Lambda,\Lambda'}\sum_{q=-1}^{1}(-1)^{J-m_j+J-\Omega}
\sqrt{(2J+1)(2J'+1)}\left(\begin{array}{ccc}
J & 1 & J'\\
-m_j & 0 & m_j
\end{array}\right)\\
&\times\left(\begin{array}{ccc}
J & 1 & J'\\
-\Omega & q & \Omega'
\end{array}\right)
\biggl\{g_L'\Lambda\delta_{\Sigma,\Sigma'}
+g_S(-1)^{S-\Sigma}\sqrt{S(S+1)(2S+1)}\\
&\times\left(\begin{array}{ccc}
S & 1 & S\\
-\Sigma & q & \Sigma'
\end{array}\right)
\biggr\}-\mu_BB_Z\sum_{q=\pm1}\delta_{\Lambda',\Lambda\pm2}(-1)^{J-m_j}
\sqrt{(2J+1)(2J'+1)}\\
&\times \left(\begin{array}{ccc}
J & 1 & J'\\
-m_j & 0 & m_j
\end{array}\right)\biggl\lbrack
(g_l'-g_{r}^{e'})(-1)^{S-\Sigma}
\left(\begin{array}{ccc}
S & 1 & S\\
-\Sigma & q & \Sigma'
\end{array}\right)\sqrt{S(S+1)(2S+1)}\\
&\times(-1)^{J-\Omega}
\left(\begin{array}{ccc}
J & 1 & J'\\
-\Omega & -q & \Omega'
\end{array}\right)-g_{r}^{e'}\delta_{\Sigma,\Sigma'}(-1)^{J-\Omega}
\sum_{\Omega''} \frac{1}{2}\biggl\lbrace(-1)^{J-\Omega''}\\
&\times\left(\begin{array}{ccc}
J & 1 & J\\
-\Omega & -q & \Omega''
\end{array}\right)
\left(\begin{array}{ccc}
J & 1 & J'\\
-\Omega'' & -q & \Omega'
\end{array}\right)\sqrt{J(J+1)(2J+1)}+(-1)^{J'-\Omega''}\\
&\times\left(\begin{array}{ccc}
J & 1 & J'\\
-\Omega & -q & \Omega''
\end{array}\right)
\left(\begin{array}{ccc}
J' & 1 & J'\\
-\Omega'' & -q & \Omega'
\end{array}\right)
\sqrt{J'(J'+1)(2J'+1)}
\biggr\rbrace
\biggr\rbrack,
\end{split}
\end{equation} 
where $\mu_B$ is the Bohr magneton, $B_z$ an externally applied magnetic field, $g_L'$ is the orbital g-factor and $g_S$ is the g-factor for the electron spin. The factors $g_l'$ and $g_{r}^{e'}$ can be computed from the lambda-doubling parameters to be \cite{devlin_measurements_2015} $g_l'=p/(2B_0)$ and $g_{r}^{e'}=-q/B_0$. Values for all the necessary parameters in the case of CaF are listed in table \ref{CaFConstantsTable}.

\begin{table}
    \centering
	\begin{tabular}{ P{3cm} | P{3cm} | P{3cm} }
		
		\hline
		& $X\Sigma^2_{1/2}$ & $A^2\Pi_{1/2}$ \\
		\hline
		\hline
		
		$A$ ($\mathrm{cm^{-1}}$) & - & $72.61743$\\
		$B_0$ ($\mathrm{cm^{-1}}$) & $0.34347$ & $0.347395$\\
		$\gamma$ ($\mathrm{cm^{-1}}$) & $0.001322$ & -\\
		$q$ ($\mathrm{cm^{-1}}$) & - & $-0.0002916$\\
		$p$ ($\mathrm{cm^{-1}}$) & - & $-0.0446$\\
		$b$ (MHz) & $109.2$ & $10.0$\\
		$c$ (MHz) & $40.1$ & $-18.25$\\
		$g_L'$ & $1$ & $1$\\
		$g_S$ & $2.0023$ & $2.0023$\\
		\hline
		
	\end{tabular}
	\caption{Molecular constants used for the $X^2\Sigma_{1/2}$ and $A^2\Pi_{1/2}$ states of CaF. }
	\label{CaFConstantsTable}
\end{table}

\begin{figure}
\begin{center}
	\includegraphics[scale=1,viewport=0 0 360 580]{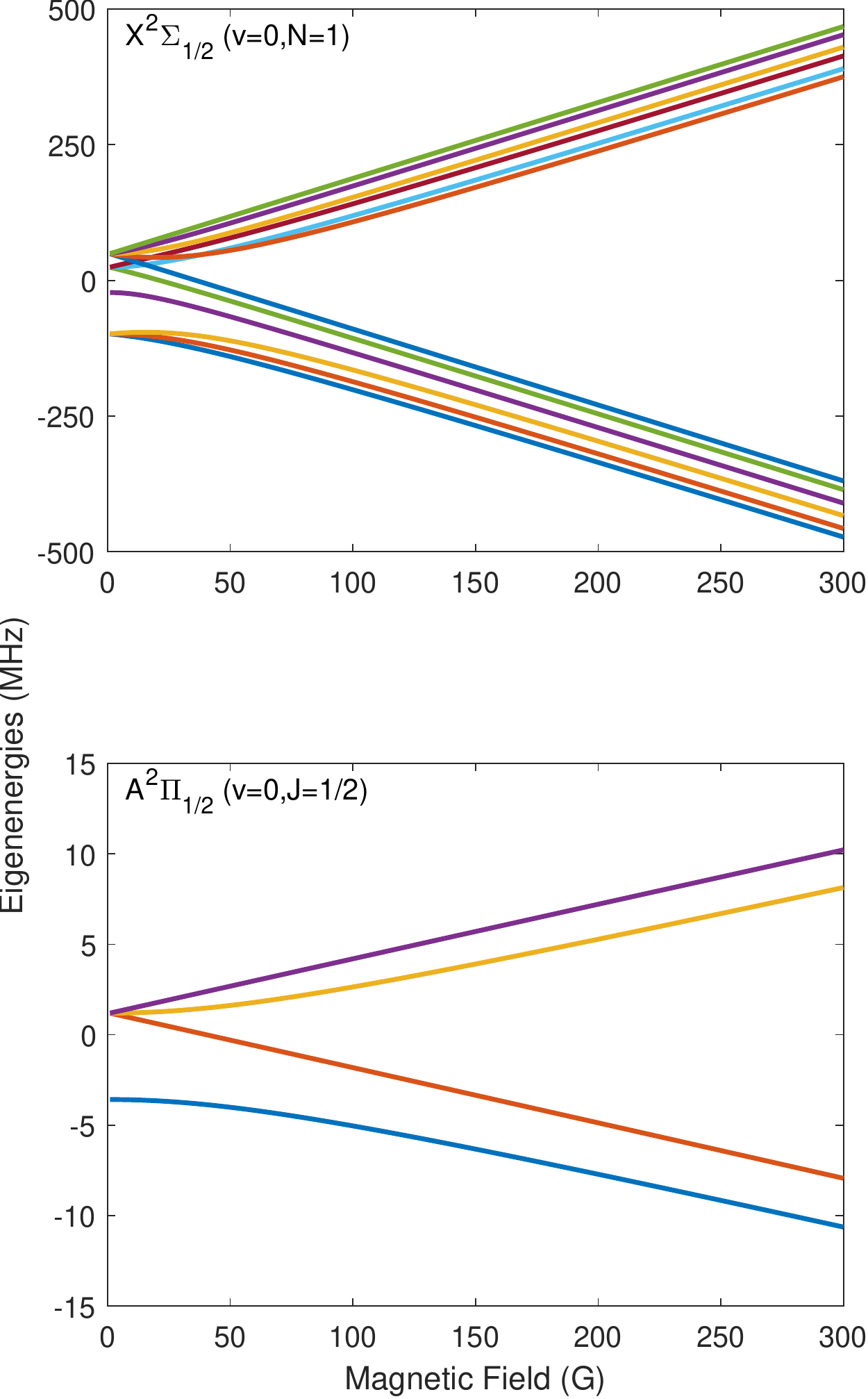}
	\caption{Eigenenergies of the $ X^2\Sigma_{1/2} (v=0, N=1)$ (upper plot) and $A^2\Pi_{1/2} (v=0, J=\frac{1}{2})$ (lower plot) manifolds of CaF, found by diagonalizing the molecular Hamiltonian given in \ref{Hamiltonian}
	}
	\label{ZeemanShifts}
\end{center}
\end{figure}

To calculate the Eigenstates and Eigenenergies at the magnetic fields of interest to us we diagonalize the Hamiltonian in \ref{Hamiltonian}. The energies vs. magnetic field of the relevant ground state ($N=1$) and excited state ($J=\frac{1}{2}$) levels are plotted in figure \ref{ZeemanShifts}.

To find the relative strengths of the optical transitions, we calculate the transition dipole matrix elements between final and initial eigenstates $\langle f\lvert \hat{d}_p \rvert i \rangle$, where $p$ denotes the polarization of the laser. Following \cite{wall_lifetime_2008} and \cite{asensio_ramos_theory_2006} we can write the matrix elements in the uncoupled basis using the Wigner-Eckart theorem as:

\begin{equation}\label{dipole_matrix_elements}
\begin{split}
\langle \Lambda&,S,\Sigma,\Omega,J,m_j,I,m_i \lvert \hat{d}_p\rvert \Lambda',S,\Sigma',\Omega',J',m'_j,I,m_i \rangle \\
&=(-1)^{J-m_j}\left(\begin{array}{ccc}
J & 1 & J'\\
-m_j & p & m'_j
\end{array}
\right) \langle \Lambda,S,\Sigma,\Omega,J,I \lvert\lvert \hat{\boldsymbol{d}}\rvert\rvert \Lambda',S,\Sigma',\Omega',J',I \rangle.
\end{split}
\end{equation}

The last term on the right hand side can be further simplified to 
\begin{equation} \label{relative_transition_strenths}
\begin{split}
\langle \Lambda&,S,\Sigma,\Omega,J,I \lvert\lvert \hat{\boldsymbol{d}}\rvert\rvert \Lambda',S,\Sigma',\Omega',J',I \rangle \\
&=\sum_{q=-1}^{1}(-1)^{J-\Omega}\sqrt{(2J+1)(2J'+1)}\left(\begin{array}{ccc}
J & 1 & J'\\
-\Omega & q & \Omega'
\end{array}
\right) \delta_{\Sigma,\Sigma'} \langle \Lambda,S,\Sigma \lvert\lvert \hat{\boldsymbol{d}}\rvert\rvert \Lambda',S,\Sigma' \rangle,
\end{split}
\end{equation}
where we have made use of the fact that the dipole operator does not couple to the spin degrees of freedom and thus $\Sigma=\Sigma'$. Since the last term in \ref{relative_transition_strenths} is the same for all transitions, we can calculate the relative transition strengths between eigenstates of $\hat{H}$ without further simplification. Table \ref{trans_prob_CaF} shows the transition strengths at high magnetic fields (Paschen-Back regime) between the states in figure \ref{ZeemanShifts}.

\begin{table}
	\begin{tabular}{ P{3cm} | P{3cm} | P{3cm} | P{3cm}| P{3cm} }
		
		\hline
		\diagbox[width=3.3cm, height=1.5cm]{$X^2\Sigma_{1/2}$}{$A^2\Pi_{1/2}$}& $m_j=\frac{1}{2},m_i=\frac{1}{2}$ & $m_j=\frac{1}{2},m_i=-\frac{1}{2}$ & $m_j=-\frac{1}{2},m_i=-\frac{1}{2}$ & $m_j=-\frac{1}{2},m_i=\frac{1}{2}$\\
		\hline
		\hline
		
		$m_s=-\frac{1}{2},m_i=\frac{1}{2},m_N=1$ &\textcolor{olive}{0.33}	&0	&0	&\textcolor{blue}{0.17}\\
		\hline
		$m_s=-\frac{1}{2},m_i=\frac{1}{2},m_N=0$ & \textcolor{red}{0.34}	&0	&0	&0 \\
		\hline
		$m_s=-\frac{1}{2},m_i=\frac{1}{2},m_N=-1$ & 0	&0	&0	&\textcolor{red}{0.17} \\
		\hline
		$m_s=-\frac{1}{2},m_i=-\frac{1}{2},m_N=1$ & 0	&\textcolor{olive}{0.33}	&\textcolor{blue}{0.17}	&0 \\
		\hline
		$m_s=-\frac{1}{2},m_i=-\frac{1}{2},m_N=0$ & 0	&\textcolor{red}{0.34}	&0	&0 \\
		\hline
		$m_s=-\frac{1}{2},m_i=-\frac{1}{2},m_N=-1$ & 0	&0	&\textcolor{red}{0.17}	&0 \\
		\hline
		$m_s=\frac{1}{2},m_i=-\frac{1}{2},m_N=-1$ & 0	&\textcolor{red}{0.16}	&\textcolor{olive}{0.33}	&0\\
		\hline
		$m_s=\frac{1}{2},m_i=-\frac{1}{2},m_N=0$ & 0	&0	&\textcolor{blue}{0.33}	&0 \\
		\hline
		$m_s=\frac{1}{2},m_i=-\frac{1}{2},m_N=1$ &0	&\textcolor{blue}{0.17}	&0	&0\\
		\hline
		$m_s=\frac{1}{2},m_i=\frac{1}{2},m_N=-1$ & \textcolor{red}{0.16}	&0	&0	&\textcolor{olive}{0.33} \\
		\hline
		$m_s=\frac{1}{2},m_i=\frac{1}{2},m_N=0$ & 0	&0	&0	&\textcolor{blue}{0.33}\\
		\hline
		$m_s=\frac{1}{2},m_i=\frac{1}{2},m_N=1$ & \textcolor{blue}{0.17}	&0	&0	&0\\
		
		\hline
		
	\end{tabular}
	\caption{Transition strengths between the $A^2\Pi_{1/2} (v=0, J=\frac{1}{2})$ and $X^2\Sigma_{1/2} (v=0, N=1)$ states of $^{40}\textrm{Ca}^{19}\textrm{F}$ at a magnetic field of \SI{1000}{\Gauss}. Strengths under 0.003 are set to zero for visual clarity. Transitions driven by $\sigma^-$-polarized light are colored blue, those driven by $\pi$-polarized light are colored brown, and $\sigma^+$ transition are denoted in red. Note that the normalization convention (sum of all decays from each upper level is 1) is slightly broken due to rounding errors.
	}
	\label{trans_prob_CaF}
\end{table}

Transition strengths below $0.003$ are set to zero for the purpose of readability. The ground states have $m_s$, $m_i$ and $m_N$ as good quantum numbers, whereas the excited states are better described by $m_j$ and $m_i$. Two important conclusions can be drawn by looking at the Transition strengths: On one hand it is possible to couple every ground state level to the excited states using only $\sigma^+$ and $\sigma^-$ light. On the other hand, in some instances two ground state levels need to be coupled to the same excited state, which may lead to the formation of coherent dark states. The fact that only $\sigma^+$ and $\sigma^-$ light are needed, allows us to build a longitudinal Zeeman slower, where slowing light travels along the magnetic field direction. In the experiment, when modulating the slowing laser at \SI{12}{\mega\hertz} to a width of \SI{96}{\mega\hertz} we see no evidence of coherent dark states lowering the radiation pressure force.

\end{document}